\documentclass[sigconf,screen, 9pt]{acmart}
\usepackage{subcaption}
\acmSubmissionID{54}
\usepackage{url}
\usepackage{enumitem}
\usepackage{graphicx}
\usepackage{caption}
\usepackage{makecell}
\usepackage{amsmath}
\usepackage{tabu}
\usepackage{gensymb}
\usepackage{multirow}
\usepackage{hyperref}
\usepackage{adjustbox}
\usepackage{amsfonts}
\usepackage{algpseudocode}
\usepackage{pgfplots}
\pgfplotsset{compat=1.12}
\usepackage{soul}
\usepackage{footnote}
\usepackage{lipsum}
\usepackage[ruled,vlined, linesnumbered]{algorithm2e}
\usepackage{subcaption}
\usepackage{listings}
\colorlet{punct}{red!60!black}
\definecolor{background}{HTML}{FFFFFF}
\definecolor{delim}{RGB}{20,105,176}
\colorlet{numb}{magenta!60!black}

\newcommand{\shrinkspace}{\vspace{-5mm}}

\newcommand*\circled[1]{\tikz[baseline=(char.base)]{
            \node[shape=circle,draw,inner sep=0.75pt, text=white,fill=black] (char) {#1};}}
\lstdefinelanguage{json}{
    basicstyle=\footnotesize\ttfamily,
    numbers=left,
    numberstyle=\scriptsize,
    xleftmargin=2.3em,
    xrightmargin=0.5em,
    framexleftmargin=1.9em,
    stepnumber=1,
    numbersep=8pt,
    showstringspaces=false,
    breaklines=true,
    frame=single,
    backgroundcolor=\color{background},
    literate=
     *{0}{{{\color{numb}0}}}{1}
      {1}{{{\color{numb}1}}}{1}
      {2}{{{\color{numb}2}}}{1}
      {3}{{{\color{numb}3}}}{1}
      {4}{{{\color{numb}4}}}{1}
      {5}{{{\color{numb}5}}}{1}
      {6}{{{\color{numb}6}}}{1}
      {7}{{{\color{numb}7}}}{1}
      {8}{{{\color{numb}8}}}{1}
      {9}{{{\color{numb}9}}}{1}
      {:}{{{\color{punct}{:}}}}{1}
      {,}{{{\color{punct}{,}}}}{1}
      {\{}{{{\color{delim}{\{}}}}{1}
      {\}}{{{\color{delim}{\}}}}}{1}
      {[}{{{\color{delim}{[}}}}{1}
      {]}{{{\color{delim}{]}}}}{1},
}

\DeclareRobustCommand*{\IEEEauthorrefmark}[1]{\raisebox{0pt}[0pt][0pt]{\textsuperscript{\footnotesize\ensuremath{\ifcase#1\or *\or \dagger\or \ddagger\or%
    \mathsection\or \mathparagraph\or \|\or **\or \dagger\dagger%
    \or \ddagger\ddagger \else\textsuperscript{\expandafter\romannumeral#1}\fi}}}}

\AtBeginDocument{%
  \providecommand\BibTeX{{%
    \normalfont B\kern-0.5em{\scshape i\kern-0.25em b}\kern-0.8em\TeX}}}

\copyrightyear{2023}
\acmYear{2023}
\setcopyright{acmlicensed}\acmConference[WoSC '23]{9th International Workshop on Serverless Computing}{December 11--15, 2023}{Bologna, Italy}
\acmBooktitle{9th International Workshop on Serverless Computing (WoSC '23), December 11--15, 2023, Bologna, Italy}
\acmPrice{15.00}
\acmDOI{10.1145/3631295.3631396}
\acmISBN{979-8-4007-0455-0/23/12}

\begin{document}

\title{GreenCourier: Carbon-Aware Scheduling for Serverless Functions}

\author{Mohak Chadha$^{1}$, Thandayuthapani Subramanian$^{1}$, Eishi Arima$^{1}$, Michael Gerndt$^{1}$, Martin Schulz$^{1}$, Osama Abboud$^{2}$
}
\affiliation{%
\institution{$^1$\{firstname.lastname, martin.w.j.schulz\}@tum.de, Technische Universit{\"a}t M{\"u}nchen 
\country{Germany}}
}
\affiliation{%
\institution{$^2$\{firstname.lastname\}@huawei.com, Huawei Technologies
\country{Germany}}
}

\renewcommand{\authors}{Mohak Chadha, Thandayuthapani Subramanian, Eishi Arima, Michael Gerndt, Martin Schulz, Osama Abboud}

\renewcommand{\shortauthors}{Mohak Chadha et al.}

\begin{abstract}
This paper presents \texttt{GreenCourier}, a novel scheduling framework that enables the runtime scheduling of serverless functions across geographically distributed regions based on their carbon efficiencies. Our framework incorporates an intelligent scheduling strategy for Kubernetes and supports Knative as the serverless platform. To obtain real-time carbon information for different geographical regions, our framework supports multiple marginal carbon emissions sources such as WattTime and the Carbon-aware SDK. We comprehensively evaluate the performance of our framework using the Google Kubernetes Engine and production serverless function traces for scheduling functions across Spain, France, Belgium, and the Netherlands. Results from our experiments show that compared to other approaches, \texttt{GreenCourier} reduces carbon emissions per function invocation by an average of $13.25$\%.

\end{abstract}



\maketitle
\vspace{-2mm}

\section{Introduction}
\label{sec:intro}
The combustion of fossil fuels for the production of electricity is a prominent driver of global greenhouse gas emissions, which has far-reaching implications for climate change on a planetary scale~\cite{fossilfuels}. According to a recent report~\cite{ipcc_report} from the U.N. Intergovernmental Panel on Climate Change, the world is rapidly approaching the Paris agreement goal~\cite{parisagreement} of limiting the Earth's average temperature increase to 1.5\degree C by the end of the 21st century. Exceeding this limit for a sustained duration can lead to irreversible damage to the environment and an increased probability of climate catastrophes~\cite{climategoal}. A significant amount of the world's total electricity is consumed by datacenters today. For instance, datacenters worldwide consumed $205$ TWh of electricity in 2018~\cite{masanet2020recalibrating}, exceeding the annual electricity consumption of countries such as Ireland and Luxembourg~\cite{owidenergy}. With the increasing migration rate of software to the cloud and the additional demand from novel domains such as Internet of Things (IoT)~\cite{wiesner2022cucumber, castro2019rise}, these datacenters are forecasted to consume between $3$\% to $13$\% of the world's total electricity by 2030~\cite{andrae2015global}.


To offset the carbon footprint from operating these datacenters, most technology companies have invested in the generation of renewable energy through sources such as wind and solar~\cite{google_sust, microsoft_sust}. For instance, Google has publicly announced its intention to achieve carbon neutrality by 2030~\cite{google_sust}. For every one MWh of renewable energy generated, these companies are awarded renewable energy credits (RECs)~\cite{acun2023carbon}. Following this, the issued RECs are used by these companies to claim \emph{net-zero} carbon emissions annually. However, due to the fluctuating nature of renewable energy~\cite{wiesner2021let},i.e., unreliability and non-availability in unlimited amounts at a single geographical location at all times, these datacenters continue to consume carbon-intensive energy repeatedly during the day when the energy supply from carbon-free sources is insufficient. According to a recent study~\cite{patros2021toward}, 50\% of the energy used in datacenters is by idle resources. To this end, an emerging cloud computing paradigm called "Serverless Computing" has the potential to reduce wasted energy consumption from resource idling and resource underutilization in datacenters.

Function-as-a-Service (FaaS) is the computational concept of serverless computing and has gained significant popularity and widespread adoption in various application domains such as machine learning~\cite{fedless, fedlesscan, serverlessfl, fastgshare}, edge computing~\cite{fado}, and heterogeneous computing~\cite{fncapacitor, jindal2021function, postericdcs}. In FaaS,  developers implement fine-grained pieces of code called \textit{functions} that are packaged independently in containers and uploaded to a FaaS platform. Several open-source and commercial FaaS platforms such as Knative~\cite{knative} and Google Cloud Functions (GCF)~\cite{gcloud-functions-2} are currently available. FaaS enables developers to focus only on the application logic while all responsibilities around resource provisioning, scaling, and related management tasks are automatically handled by the cloud provider. In addition, the FaaS computing model offers several advantages such as rapid scalability during request bursts, automatic scaling to zero when resources are unused, and an attractive pricing and development model~\cite{castro2019rise}.



Serverless functions are \textit{ephemeral}, i.e., short-lived, and \textit{event-driven}, i.e., these functions only get executed in response to external triggers such as HTTP requests. Moreover, these functions are \textit{stateless}, i.e., any application state needs to be persisted in external storage  or databases. For commercial FaaS providers, the functions are executed in the cloud providers' traditional IaaS virtual machine (VM) offerings hosted in their internal geographically distributed datacenters~\cite{chadha2021architecture, demystifying}. Typically, these cloud providers receive around one billion function invocations each day~\cite{shahrad2020serverless}.  For all commercial FaaS platforms, the geographical regions for function execution are pre-selected by the users during function deployment. As a result, the varying \textit{carbon intensities} of the different geographical regions are not considered for function execution by any commercial FaaS platform. The carbon intensity of a geographical region describes the amount of carbon emitted per kWh of energy produced. The intensity can vary significantly across regions and even within the same region over time~\cite{wiesner2021let}. Towards sustainable serverless computing, our key contributions are:




\vspace{-1mm}
\begin{itemize}
        \item We implement and present \texttt{GreenCourier}, a novel scheduling framework that intelligently schedules serverless functions across geographically distributed Kubernetes clusters to minimize carbon emissions for function execution at runtime. Our framework is available at: \url{https://github.com/kky-fury/carbon-sched}.
        \item We comprehensively evaluate our framework using the Google Cloud platform and production function traces~\cite{shahrad2020serverless} on different serverless functions against other scheduling strategies wrt generated carbon emissions and function response times. 
        \item We quantify and analyze the overhead for function scheduling in our framework. 

\end{itemize}

The rest of the paper is structured as follows. \S\ref{sec:greencourier} describes our scheduling framwork, i.e., \texttt{GreenCourier} in detail. In \S\ref{sec:expresults}, our experimental results are presented. \S\ref{sec:relatedwork} describes some previous approaches related to our work. Finally, \S\ref{sec:conclusion} concludes the paper and presents an outlook.

\vspace{-3mm}
\section{GreenCourier}
\label{sec:greencourier}

\begin{figure}[t]
    \centering
    \includegraphics[width=0.9\columnwidth]{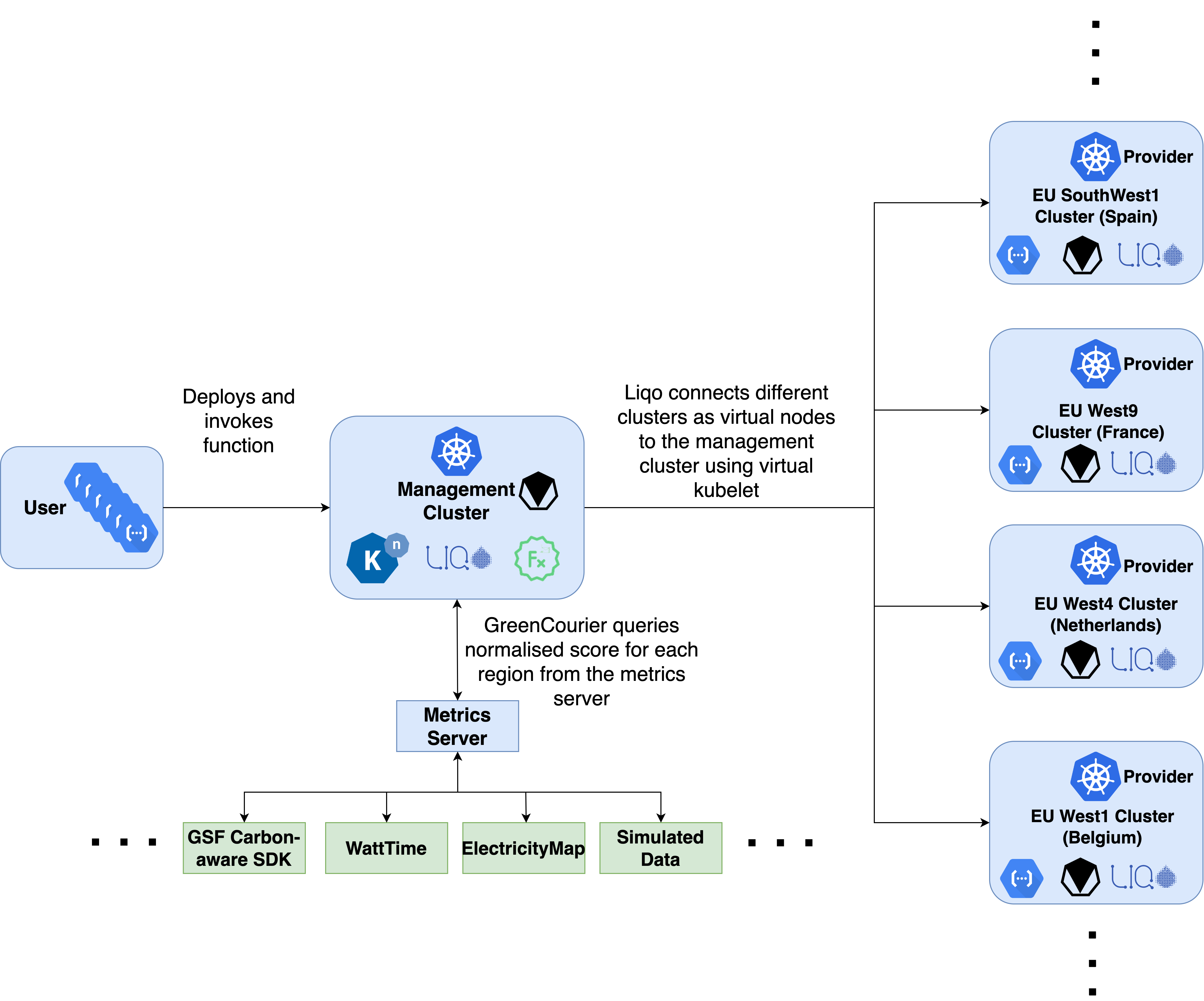}
    \vspace{-2mm}
    \caption{\texttt{GreenCourier} System Architecture.}
    \label{fig:sys_arch}
\shrinkspace
\vspace{-2mm}
\end{figure}

 \subsection{Overview}
\label{sec:overview}
The purpose of \texttt{GreenCourier} is to dynamically schedule serverless functions across different geographical regions depending on their carbon efficiency. The different system components of our framework are shown in Figure~\ref{fig:sys_arch}. GreenCourier builds on Kubernetes (K8s) which is the \emph{de-facto} platform for container orchestration in the cloud. As the FaaS platform, our framework utilizes the serving component of Knative. We chose Knative since it is utilized in production by Google in their FaaS and serverless Container-as-a-Service offerings~\cite{gcloud-functions-2,GCR}. For establishing geographically distributed multi-cluster K8s topologies, our framework uses Liqo~\cite{liqo}. In multi-cluster K8s environments, independent clusters can be located in different datacenters across different countries leading to several challenges such as synchronization of cluster objects for pod \footnote{We use the term pod and function interchangeably in this paper since they are the same from K8s's perspective.} scheduling and communication between services in different clusters. Liqo is an open-source project that provides a multi-cluster control plane that enables (i) the seamless scheduling of pods across different clusters and (ii) the automatic dynamic discovery of new clusters as they are added to the environment. In addition, it provides a network fabric~\cite{liqo_network}, which extends the K8s networking model across all clusters in the environment, enabling the inter-communication between services. For connecting two individual clusters, Liqo uses the \emph{peering} methodology~\cite{liqo_peering}. Peering involves establishing a unidirectional resource and service consumption relationship between two clusters, where one cluster, i.e., the consumer is granted the ability to offload pods to another remote cluster, i.e., the provider, but not vice-versa. In this scenario, the consumer creates an outgoing peering connection towards the provider, while the provider is subjected to an incoming peering connection from the consumer. This allows for the two clusters to interconnect and enables the consumer to access the services and resources provided by the provider cluster. In our framework, the different geographically distributed clusters are the providers, while the management cluster is the consumer. For connecting the independent provider clusters to the management cluster, Liqo uses Virtual Kubelet (VK)~\cite{virtual_kubelet}. VK cloaks a provider cluster as a virtual node and connects it to the management cluster. For carbon-aware scheduling decisions at runtime, our framework obtains normalized carbon score data from the metrics server (\S\ref{sec:metricsserver}). Following this, the function is scheduled to the region with the highest current carbon score. The management cluster is the central component of our framework, where the \texttt{GreenCourier} scheduler (\S\ref{sec:workflow}), Knative, Liqo, VK, and the metrics server are deployed. From the perspective of the user, using our framework is extremely uncomplicated. A user simply deploys their function written in any programming language supported by Knative on our management cluster. Following this, the function can be invoked using the returned \texttt{url} after function deployment.

\vspace{-3mm}

\subsection{Metrics Server}
\label{sec:metricsserver}

The metrics server is responsible for calculating and normalizing the carbon-efficiency scores for the different geographical regions. It exposes a REST API that can be efficiently utilized by the \texttt{GreenCourier} scheduler (\S\ref{sec:overview}) to obtain the normalized carbon intensity scores. To obtain data about carbon emissions, our metrics server currently supports two sources, i.e., WattTime (default)~\cite{watt_time} and the Carbon-aware SDK~\cite{carbon_sdk}. WattTime leverages real-time grid data and advanced algorithmic and forecasting techniques to present unprecedented visibility into the marginal operating emission rate (MOER) of the local electricity grid. MOER is a quantification of the rate of carbon emissions produced by  electricity generators in response to variations in the demand for energy on the local electrical grid, at a specific point in time and geographical location. WattTime provides energy data in units of pounds of emissions per megawatt-hour, i.e., lbsCO2/MWh. The Carbon-aware SDK was developed by the Green Software Foundation (GSF) to enable the development of advanced software solutions that embody a carbon-conscious approach to energy consumption. Towards this, it provides a REST API and command line interface that can be utilized by carbon-aware applications to obtain aggregated MOER data from different third-party sources such as WattTime. In addition, it provides a standardized interface for obtaining MOER data in units of grams of emissions per kilowatt-hour, i.e., gCO2/kWh. For normalizing the current carbon scores, we use the \emph{min-max} normalization technique in our metrics server. Furthermore, our metrics server can be easily extended to support other carbon data sources such as Electricity Maps~\cite{electricityMap} and simulated data~\cite{wiesner2021let}.

\setlength{\textfloatsep}{0pt}
\begin{algorithm}[t]
\captionsetup{skip=0pt}

\SetAlgoLined
\SetKwFunction{train}{Client\_Update}
\SetKwFunction{calcSccores}{Calculate\_Scores}
\SetKwFunction{retriveData}{Start\_Timer}
\SetKwFunction{stopTime}{Stop\_Timer}
\SetKwProg{Pn}{Function}{:}{\Return{$PodObject$}}
\DontPrintSemicolon
\textbf{Scoring Regions:}
\;

\Pn{\calcSccores{$PodObject$, $ListofNodes$}}
{   
    \For{each $Node$ in $ListofNodes$}
    {
       $Region$ = Node.Annotation("region").\;
       $NodeScore$ = Retrieve carbon score from metrics server for $Region$.\;
       Update and store $NodeScore$.\;
    }
    Normalise node scores.\;
    $Node$ = Get the node with the highest score.\;
    Assign $PodObject$ on $Node$.
}

\caption{Carbon-aware scheduling strategy.}
\label{alg:carbon-algo}
\end{algorithm}
\setlength{\textfloatsep}{12.0pt plus 2.0pt minus 2.0pt}

\vspace{-3mm}
\subsection{Carbon-aware scheduling of serverless functions}
\label{sec:carbonawaresched}
The carbon-aware scheduling strategy in \texttt{GreenCourier} is implemented using the Scheduling framework~\cite{scheduling_framework} provided by K8s. The scheduling framework  defines various extension points exposed by the K8s scheduler API, where different plugins can be enabled and executed during pod scheduling. Scheduling of pods in K8s occurs in two phases, i.e., the scheduling cycle and the binding cycle. The former involves selecting a node for executing the pod, while the latter involves applying that decision to the cluster. Node selection in K8s involves two consecutive stages, i.e., filtering and scoring.  The filtering phase involves the execution of a set of predicate plugins that define strict limitations on a node to be eligible for executing a pod. For instance, the \emph{NodeResourcesFit} filtering plugin checks whether the resources requested by a pod are available on a particular node. Our scheduling strategy supports multiple predicate plugins provided by K8s~\cite{scheduling_schemes} such as \emph{NodeResourcesFit}, \emph{TaintToleration}, and \emph{NodeAffinity}. In contrast, the scoring phase evaluates a node based on a series of soft constraints called priorities. These priorities may either be specified in the pod definition or could stem from broader constraints associated with the nodes or the cluster. In this phase, every node is evaluated by the enabled priority plugins, which assign every node a score. For instance, the \emph{ImageLocality} priority plugin, assigns a node with a high score if the container image requested by the pod is locally present on that node. On completion of the scoring phase, the node with the highest aggregated score is selected for pod assignment. To enable the carbon-aware scheduling of serverless functions across geographically distributed clusters, we implement a custom scoring plugin for K8s. Algorithm~\ref{alg:carbon-algo} describes our custom scoring strategy. \emph{ListofNodes} represent the eligible nodes available for pod assignment after the filtering phase, while \emph{PodObject} represents the current pod that needs to be scheduled. Initially, our scoring plugin iterates over the eligible set of nodes and obtains the current carbon score associated with the region where the node is located, from the metrics server (Lines 3-5). The region of the node is determined by reading the annotations set by the administrator during cluster creation (Line 4). We store and update the obtained carbon scores in a key-value-based data structure (Line 5). To reduce overhead for scheduling, we cache the obtained carbon scores for a particular region for five minutes locally. We chose this granularity since both WattTime and Carbon-aware SDK provide updated data in five-minute intervals (\S\ref{sec:metricsserver}). After iterating across all nodes, we normalize the obtained node scores between zero to $100$ (Line 8). Following this, we select the node with the highest score (Line 9) and assign the pod specification with its name (Line 10).

\vspace{-2mm}

\begin{figure}[t]
    \centering
    \includegraphics[width=\columnwidth]{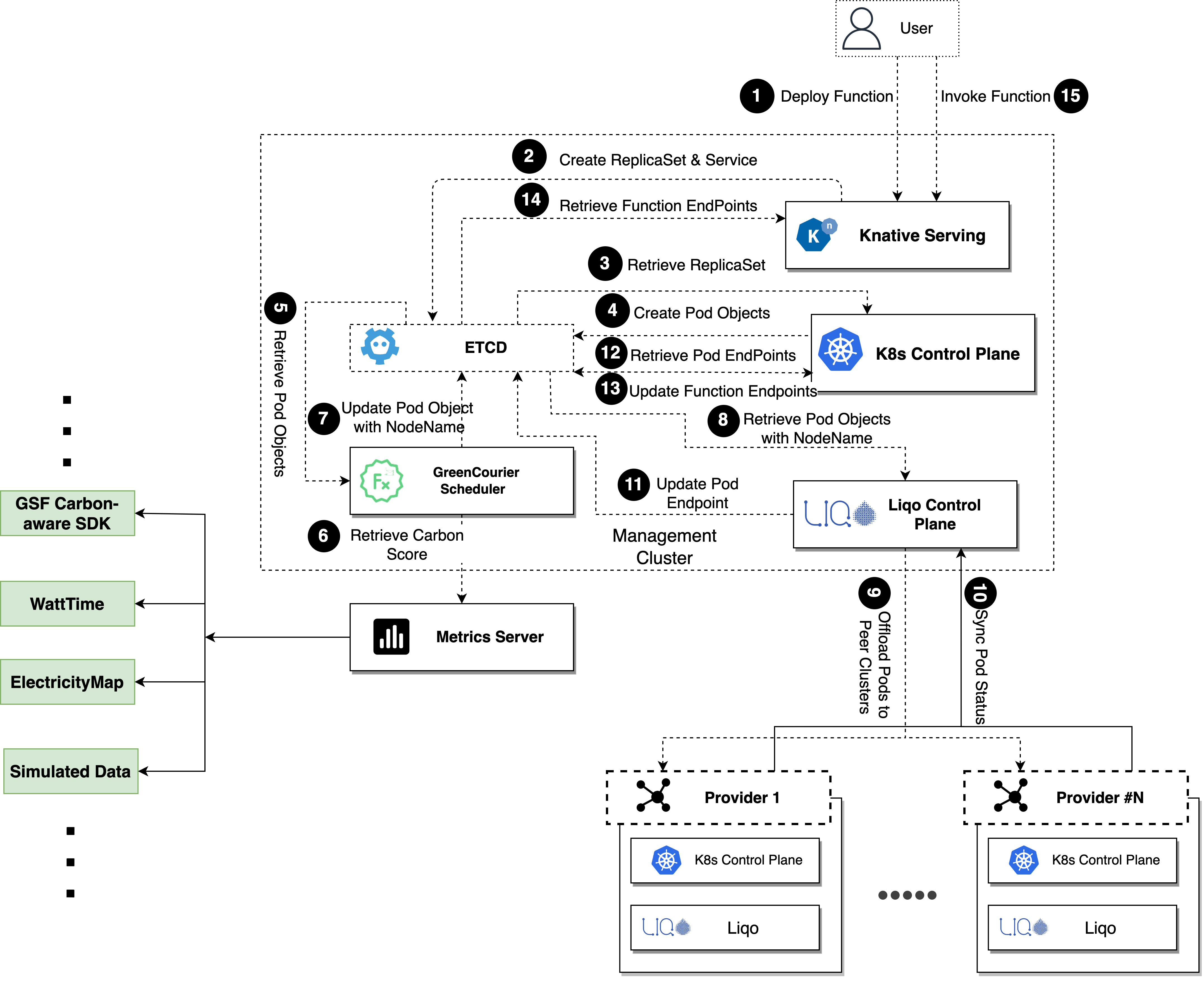}
    \caption{Workflow for serverless function scheduling in \texttt{GreenCourier}.}
    \label{fig:workflow}
\shrinkspace
\vspace{-2mm}
\end{figure}

\subsection{Putting It All Together}
\label{sec:workflow}
Figure~\ref{fig:workflow} shows the detailed workflow for scheduling serverless functions in \texttt{GreenCourier}. First, the user defines the serverless function as a \texttt{YAML} specification file that contains information regarding its container image and runtime configurations (\circled{1}). To use our carbon-aware scheduler, the user must set the field \emph{schedulerName} to \texttt{kube-green-courier} in the specification file. Using this specification file, the user sends a request to the Knative serving control plane to create the function. Following this, the associated K8s objects are created by Knative (\circled{2}). This leads to a state change in \texttt{etcd}, which is detected and retrieved by K8s (\circled{3}). After this, K8s creates the associated Pod objects for the function and updates \texttt{etcd} (\circled{4}). Our custom scheduler listens for any changes to the state associated with Pod objects for Knative functions. As a result, the newly created pod objects are retrieved by our scheduler (\circled{5}). Following this, our implemented priority plugin (\S\ref{sec:carbonawaresched}) is executed, which involves obtaining the carbon-efficiency scores for the different geographically distributed nodes (\circled{6}). On selection of the optimal node for pod assignment, our scheduler modifies the \texttt{nodeName} field in the pod specification and pushes the updated specification to \texttt{etcd} (\circled{7}). Liqo~\cite{liqo} acts as an alternative to \texttt{kubelet}~\cite{kubelet_doc} in our scheduling framework. It listens for pod objects that are assigned to the different virtual nodes connected to our management cluster and retrieves them from \texttt{etcd} (\circled{8}). After this, Liqo initiates the pod offloading process (\circled{9}) to the chosen provider cluster (\S\ref{sec:overview}). On reconciliation of the pod creation request, Liqo updates the pod's status in the management cluster (\circled{10}). Following this, Liqo updates the local routing tables of the management cluster to ensure that all network traffic to the offloaded pod is sent through the network fabric established after the peering process (\S\ref{sec:overview}). For this, it updates the endpoint, i.e., IP of the pod associated with the serverless function (\circled{11}). The updated pod endpoints are retrieved by K8s (\circled{12}) which triggers an update for the deployed Knative function (\circled{13}). Following this, Knative updates the created pod objects of the function with the appropriate IP of the pod running in the provider cluster (\circled{14}). Finally, the function can be invoked by the user through the \texttt{url} received after function deployment (\circled{15}). On multiple invocations of the deployed function, the Knative pod autoscaler (KPA) increases the replica count of the deployed function to reduce function response times. In this scenario, steps \circled{3}-\circled{14} are repeated.

\begin{table}[t]
    \centering
    \begin{adjustbox}{width=8.8cm,  center}
    \begin{tabular}{|c|c|c|c|c|c|}
    \hline
        \textbf{Cluster Type} &  \textbf{Geographical Region}  & \textbf{Instance type} & \textbf{\# of instances} & \textbf{Total \# of vCPUs} & \textbf{Total Memory (GiB)} \\ \hline
         Management & europe-west3-a (Frankfurt) & e2-standard-16 & 1 & $16$ & $64$ \\ \hline
         \multirow{4}{*}{Provider} & europe-southwest1-a (Spain) &e2-standard-4 & 4 &$ 16$ & $64$ \\ \cline{2-6}
         & europe-west9-a (France) & e2-standard-4 & 4 & $16$ & $64$\\ \cline{2-6}
         & europe-west1-b (Belgium) & e2-standard-4 & 4 & $16$ & $64$\\ \cline{2-6}
         & europe-west4-a (Netherlands) & e2-standard-4 & 4 & $16$ & $64$\\ \hline
    \end{tabular}
    \end{adjustbox}
    \caption{Cluster configurations used for evaluating \texttt{GreenCourier}.}
    \label{tab:cloud-resource}
\shrinkspace
\vspace{-4mm}
\end{table}

\vspace{-2mm}

\section{Experimental Results}
\label{sec:expresults}
In this section, we present performance results for \texttt{GreenCourier}. For all our experiments, we follow best practices while reporting results.

\subsection{Experimental Setup}
\label{sec:expsetup}

\subsubsection{Experiment configuration}
\label{sec:expconfig}

For analyzing the performance of our scheduling framework, we use the Google Kubernetes Engine (GKE) in standard mode~\cite{gke}. Table~\ref{tab:cloud-resource} shows the different clusters and their configurations used in our experimental setup. We deploy our management cluster (\S\ref{sec:overview}) in the Frankfurt region, i.e., \texttt{europe-west-3a} with a configuration of $16$vCPUs and $64$GiB of memory. In addition, we provision four provider clusters across Madrid (\texttt{europe-southwest1-a}), Paris (\texttt{europe-west9-a}), St. Ghislain (\texttt{europe-west1-b}), and Eemshaven (\texttt{europe-west4-a}). Each provider cluster has four VMs of type \texttt{e2-standard-4}. In our setup, we only consider geographical locations within Europe due to the limitations of our WattTime license for obtaining carbon data outside it. All VMs provisioned in our setup are based on the Intel Xeon Platinum 8173M Skylake-SP processor architecture~\cite{procgcp}. Note that by default GKE does not support the provisioning of a K8s cluster with nodes in different geographical regions.

\begin{table}[t]
    \centering
    \begin{adjustbox}{width=6cm,  center}
    \begin{tabular}{|c|c|}
    \hline
    \textbf{Function Name} & \textbf{Description} \\ \hline
        \makecell{CNN-Serving} &  \makecell{Image classification using the CNN Squeeze net architecture.}\\ \hline
         \makecell{Float} & \makecell{Performs a series of floating point arithmetic operations \\  such as squareroot,  sine, and cosine.} \\ \hline
         \makecell{LR-Serving} & \makecell{Score prediction using logistic regression based on reviews \\ as input for the Amazon reviews dataset.} \\ \hline
         \makecell{Linpack} & \makecell{Solves a dense n $\times$ n system \\ of linear equations.} \\ \hline
         \makecell{Matrix \\ Multiplication} & \makecell{Matrix multiplication of two square matrices.} \\ \hline
         \makecell{PyAES} & \makecell{Private key-based encryption and decryption. A pure Python \\ implementation of the AES block-cipher algorithm in CTR mode.} \\ \hline
         \makecell{RNN-Serving} & \makecell{Word Prediction using the Recurrent Neural Network architecture.} \\ \hline
         \makecell{Chameleon} & \makecell{Render a template using the  \\ \texttt{Chameleon} library to create an HTML table.} \\ \hline
    \end{tabular}
    \end{adjustbox}
    \caption{The different serverless functions used in this work.}
    \label{table:funcdesc}
\shrinkspace
\vspace{-4mm}
\end{table}

\subsubsection{Serverless functions}
\label{sec:servfuncs}

For our experiments, we use eight functions (\S\ref{sec:carbeff},\S\ref{sec:resptimes}) from the \texttt{FunctionBench}~\cite{kim2019functionbench} benchmark suite. The different functions are described in Table~\ref{table:funcdesc}. We adapt the different functions for Knative and modify the communication protocol from HTTP to \texttt{gRPC}. All functions are implemented using the Python programming language.


\subsubsection{Load Testing}
\label{sec:loadtesting}
For emulating actual user behavior for invoking the deployed serverless functions, we implement a custom load generator using \texttt{k6}. \texttt{k6} is an open-source performance and regression testing tool that provides a JavaScript API to develop custom load-testing scenarios. In addition, it supports both HTTP(s) and \texttt{gRPC} based protocols for communication. To ensure a representative function invocation pattern in our experiments, we use the production-level serverless function traces from Microsoft Azure functions~\cite{shahrad2020serverless}. Our load generator reads the input traces and generates different scenarios, dynamically simulating the function request pattern. To model request inter-arrival time, we use the Poisson distribution. For our experiments, we limit a load test to ten minutes and repeat it five times.


\begin{footnotesize}
\begin{equation}
\label{eq:sci}
    Software \; Carbon\; Intensity = ((E \cdot I) + M ) /  R
\end{equation}
\shrinkspace
\vspace{-2mm}
\end{footnotesize}

\begin{footnotesize}
\begin{equation}
\label{eq:calci}
W.\; A. \; MOER = \frac{\sum_{i=1}^{n}{(\# function\,instances\,in\,a\, region(i) \;* \; MOER(i))}}{\sum_{i=1}^{n}{\# function\, instances\,in\,a\, region(i)}}    
\end{equation}
\shrinkspace
\vspace{-2mm}
\end{footnotesize}


\begin{figure*}[t]
\centering
\begin{subfigure}{0.49\textwidth}
    \centering
        \includegraphics[width=0.49\columnwidth]{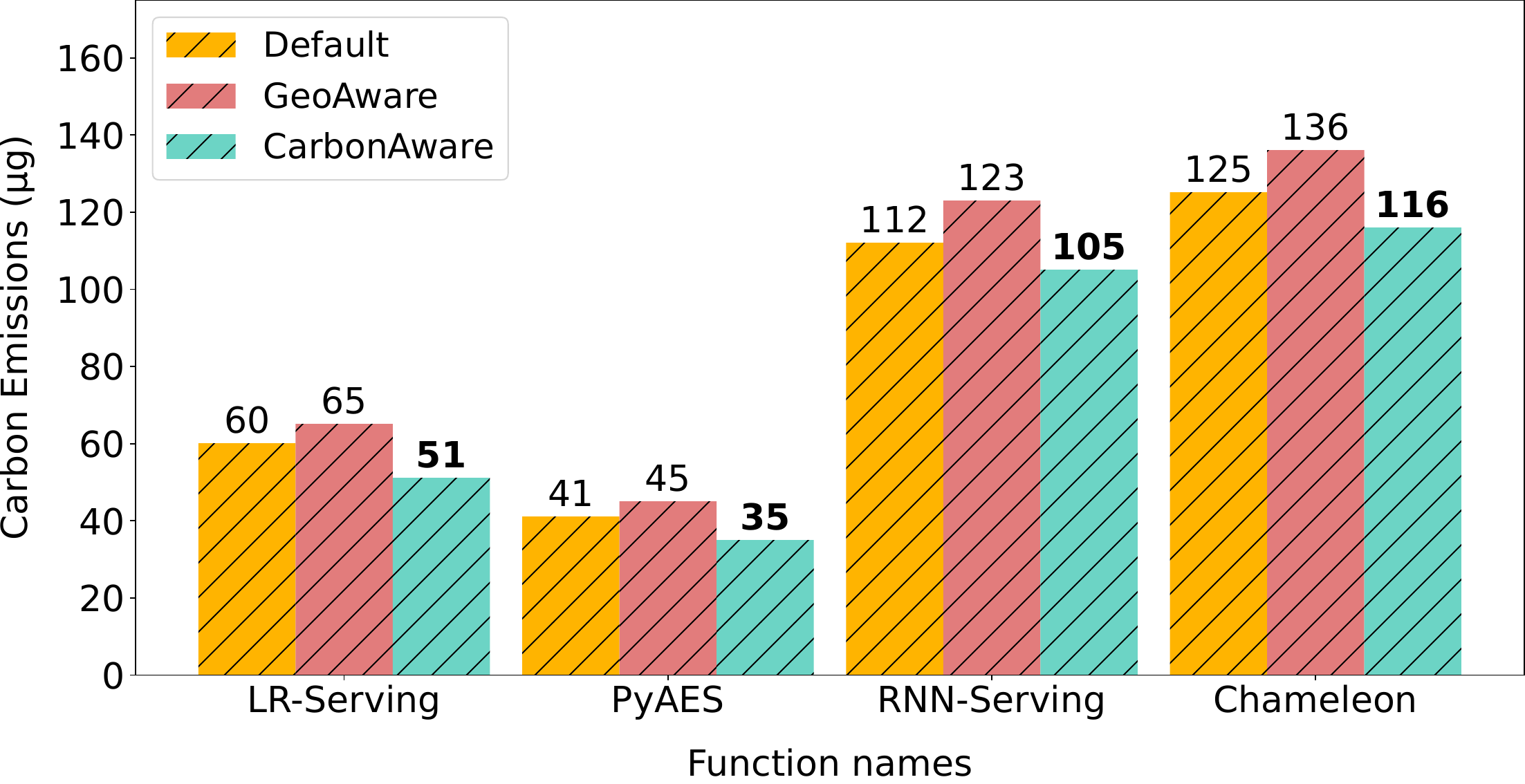}
        \includegraphics[width=0.49\columnwidth]{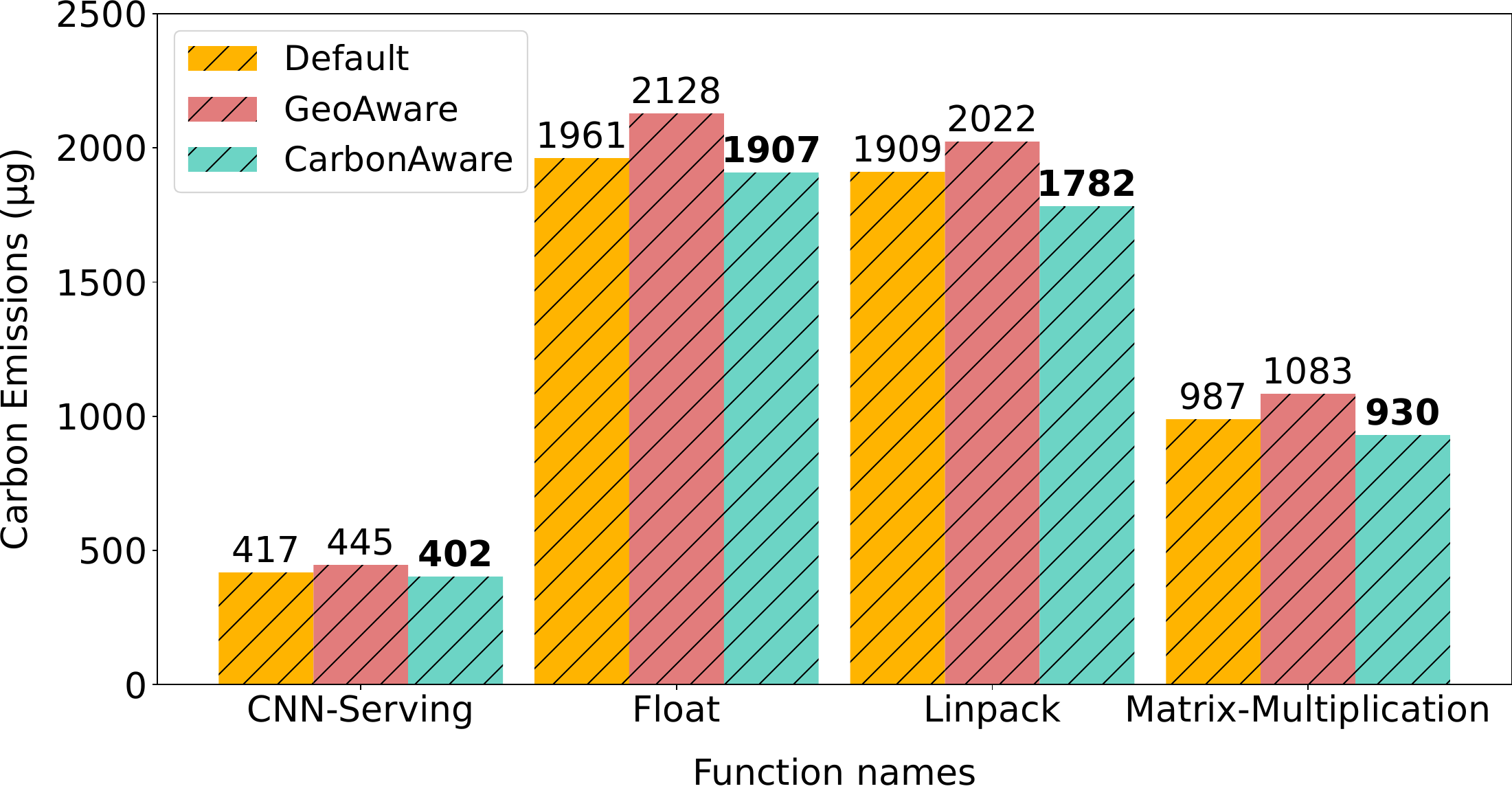}
        \caption{Comparing carbon emissions.}
        \label{fig:compcarbon}    
\end{subfigure}
\begin{subfigure}{0.49\textwidth}
    \centering
        \includegraphics[width=0.49\columnwidth]{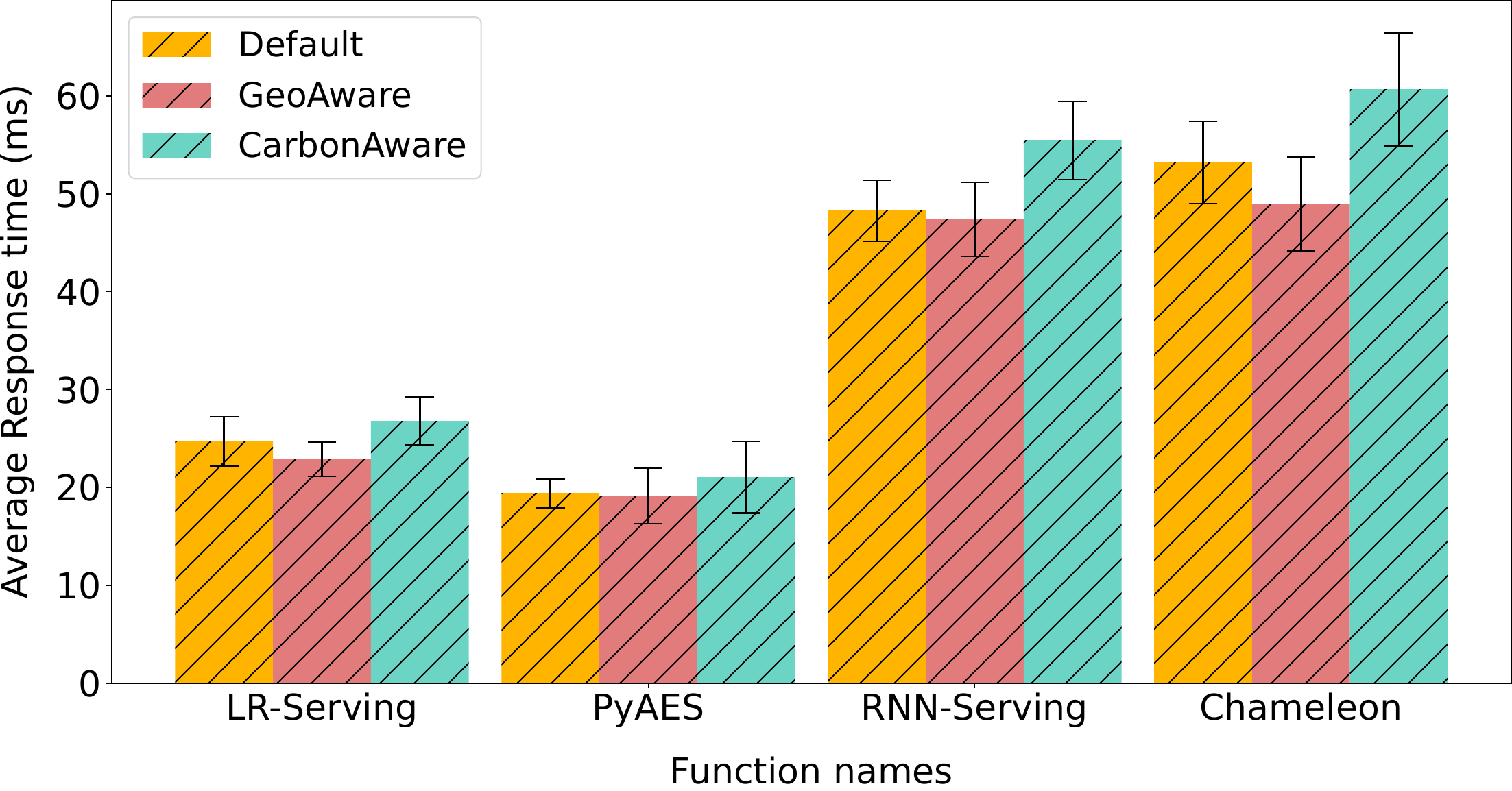}
        \includegraphics[width=0.49\columnwidth]{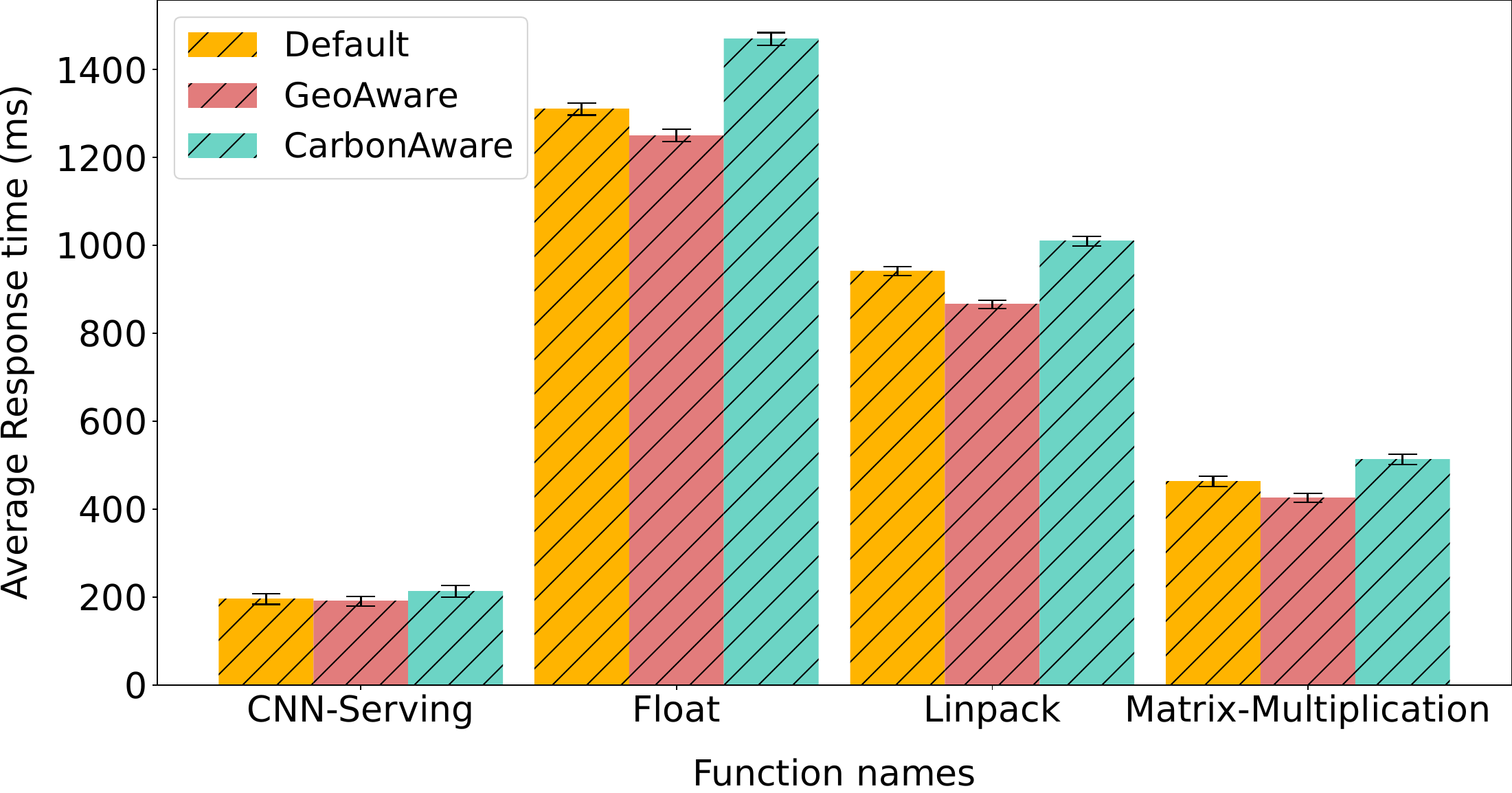}
        \caption{Comparing response times.}
        \label{fig:compperf}    
\end{subfigure}
\caption{Comparing carbon emissions and response times for the different serverless functions (\S\ref{sec:servfuncs}) and scheduling strategies.}
\label{fig:metrics}
\shrinkspace
\vspace{-2mm}
\end{figure*}

\begin{figure}[t]
\centering
\begin{subfigure}{0.9\columnwidth}
    \centering
        \includegraphics[width=0.45\columnwidth]{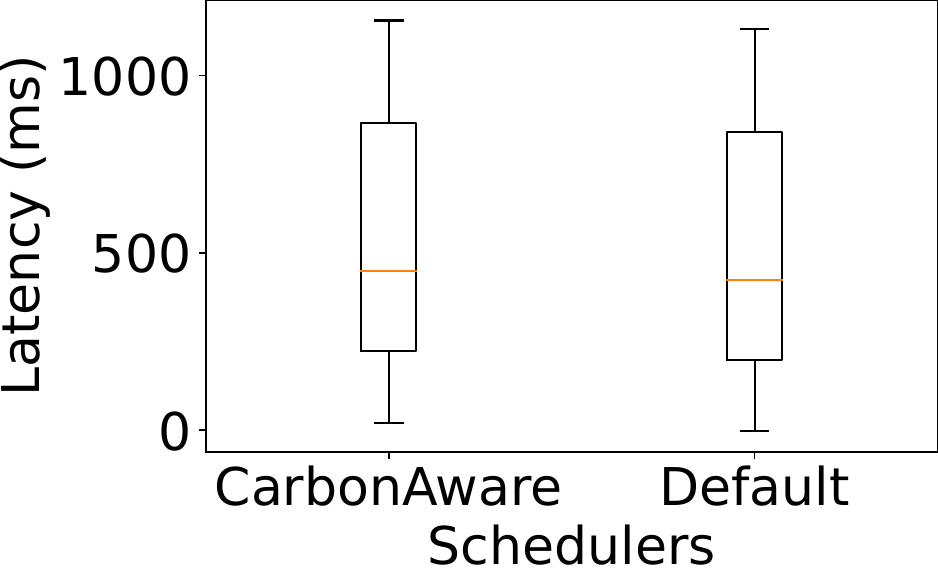}
        \includegraphics[width=0.45\columnwidth]{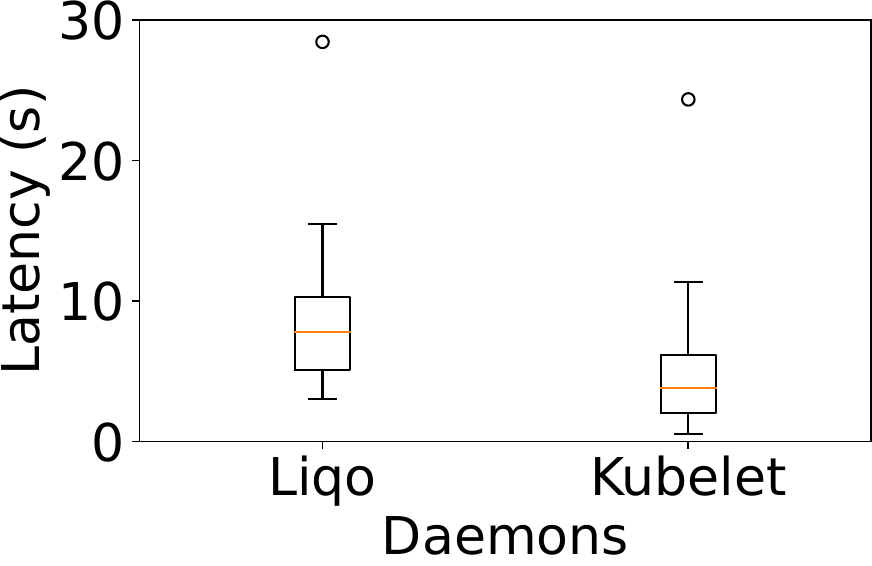}
        \label{fig:schedlatency}    
\end{subfigure}
\vspace{-2mm}
\caption{Comparing scheduling latency (left) and binding latency (right). in our scheduling framework.}
\label{fig:ovearhead}
\shrinkspace
\end{figure}

\subsubsection{Metrics}
\label{sec:metrics}
To quantify and compare the performance of our scheduling framework, 
we use the metrics \textit{carbon emissions}, \textit{function response time},  
\textit{scheduling latency}, and \textit{binding latency}. For computing the 
carbon emissions per function invocation, we use the Software Carbon Intensity 
(SCI) specification~\cite{sci} provided by the GSF (\S\ref{sec:metricsserver}). 
SCI provides a detailed guide for determining the carbon footprint of a software application 
by quantifying its associated carbon emissions. For any application, SCI can be 
calculated by using Equation~\ref{eq:sci}. In Equation~\ref{eq:sci}, 
\texttt{E} represents the energy consumed by the application, \texttt{I} 
represents the location-based marginal carbon intensity, \texttt{M} 
represents the embodied carbon emissions for the hardware where the 
application is executing, and \texttt{R} represents the functional unit 
such as the \#requests by which an application scales. To estimate \texttt{E} 
for our provider clusters  with $64$vCPUs and $256$GiB of RAM, 
we use the methodology described in~\cite{calcsci}. Towards this, 
we use $165$W as the TDP for the Intel Skylake-SP processor~\cite{tdp} and 
assume $50$\% CPU utilization. We choose this number since recent studies 
suggest that cloud computing resources are over-provisioned and not fully 
utilized~\cite{resource}. Similar to~\cite{calcsci}, we assume that every 
$8$GiB of RAM consumes $3$W of power. In addition, $2$vCPUs constitute one 
CPU core on GKE. Following this, we calculate $E$ for our cluster setup as 
$165\times50\%\times24*32+96=63.456$kWh. To compute \texttt{I}, we calculate 
the weighted average MOER as shown in Equation~\ref{eq:calci}. This is because 
the number of function instances\footnote{A single function instance is 
equivalent to a single pod in K8s.} launched after a load test can vary across 
regions and a weighted average constitutes a valid approximation. In Equation~\ref{eq:calci}, 
\texttt{n} represents the total number of regions.
For estimating \texttt{R}, we assume that the request traffic is uniformly
 distributed among all function instances. Following this, we calculate the 
 maximum number of requests that can be served by a single function instance 
 during a day using its response time. For instance, for a function with a 
 response time of $200$ms the \texttt{R} value would be $432000$ requests 
 per 24 hours. In our calculations, we ignore \texttt{M} since embodied 
 carbon emissions remain unaffected by the implementation of our scheduling 
 framework.

Function response time represents the end-to-end time to process and respond to a function invocation request. For computing the scheduling latency (\S\ref{sec:carbonawaresched}), we measure the time between the \textit{NodeAssgined} event generated by the K8s scheduler and the \textit{PodCreation} event generated by the ReplicaSet controller. On the other hand, for computing the binding latency, we measure the time between the \textit{NodeAssgined} event and the \textit{PodRunning} event generated by Liqo/Kubelet.


\vspace{-3mm}

\subsection{Comparing carbon emissions}
\label{sec:carbeff}
We compare the performance of our carbon-aware scheduling strategy against the default scheduling strategy in K8s and a custom \texttt{GeoAware} scheduling strategy. In contrast to our carbon-aware scheduling strategy, the \texttt{GeoAware} strategy assigns pods to nodes that are located at a closer distance to the management cluster. Similar to our carbon-aware strategy (\S\ref{sec:carbonawaresched}), we implement the \texttt{GeoAware} scheduling strategy as a priority plugin in K8s. Figure~\ref{fig:compcarbon} shows the carbon emissions in $\mu gs$ for a single function invocation across the different scheduling strategies. We quantify the amount of carbon emitted using the SCI specification as described in \S\ref{sec:metrics}. For the different serverless functions, using \texttt{GreenCourier} leads to $8.7$\% and $17.8$\% reduction in carbon emissions on average as compared to the default and the \texttt{GeoAware} scheduling strategies respectively. In our experimental setup (\S\ref{sec:expsetup}), the most carbon-efficient region is \texttt{europe-southwest1-a}, followed by \texttt{europe-west9-a}, \texttt{europe-west1-b}, and \texttt{europe-w\-est4-a}. On the other hand, the region closest to the management cluster is \texttt{europe-west1-b}, followed by \texttt{europe-west\-4-a}, \texttt{europe-west9-a}, and \texttt{europe-southw\-est1-a}. As a result, the \texttt{GeoAware} strategy places serverless functions in regions with low carbon efficiencies leading to a higher carbon footprint. The default scheduling strategy performs better than the \texttt{GeoAware} strategy since it relies on the \texttt{PodTopology\-Spread} K8s plugin that tries to evenly spread functions across all provider clusters to maximize availability.

While the geo-based scores for the different provider clusters always remain constant, the carbon-scores for the different regions change every five minutes (\S\ref{sec:carbonawaresched}). However, in our experiments, we observed that \texttt{europe-southwest1-a} and \texttt{europe-west9-a} were always the most carbon-efficient regions.  It is important to note that the carbon efficiencies of the different geographical regions selected for deploying the provider clusters influence the carbon reductions achieved with our scheduling strategy (\S\ref{sec:carbonawaresched}). With significant differences between the carbon scores of the different geographical regions, the reductions in carbon emissions will be significantly higher as compared to the other scheduling strategies, and vice versa.


\vspace{-3mm}

\subsection{Comparing response times}
\label{sec:resptimes}
Figure~\ref{fig:compperf} shows the average response times for the different serverless functions and scheduling strategies. From our experiments, we observe that the \texttt{GeoAware} scheduling strategy leads to the least average response times for the different serverless functions, followed by the default and the carbon-aware scheduling strategies. Our carbon-aware scheduling strategy leads to geometric mean (GM) average slowdown of $10.26$\% and $16.24$\% across all functions as compared to the default and \texttt{GeoAware} scheduling strategies respectively. As compared to the default scheduling strategy, the \texttt{GeoAware} strategy leads to a GM average speedup of $4.2$\% across all functions. The performance degradation with our carbon-aware scheduling strategy can be attributed to the selection of carbon-efficient regions, i.e., \texttt{europe-southwest1-a} and \texttt{europe-west9-a} for serverless function execution. These geographical regions are furthest away from our management cluster, i.e., (\texttt{europe-west3-a}) leading to higher average function response times. On the other hand, the \texttt{GeoAware} strategy maximizes the execution of functions in regions closest to our management cluster leading to significantly less average response times. 
Note that the cost for all three strategies is equal since the users are only billed for the function execution time in the FaaS computing model.


        
\vspace{-3mm}

\subsection{Analyzing overhead}
\label{sec:ovearhead}
Figure~\ref{fig:ovearhead} shows the comparison between the scheduling latencies for our carbon-aware scheduler and the default scheduler in K8s (\S\ref{sec:carbonawaresched},\S\ref{sec:metrics}). The yellow line represents the median observed latency, while the whiskers represent the minimum and maximum latency values. We omit the comparison with the \texttt{GeoAware} scheduling strategy since its implementation is similar to our carbon-aware strategy. From our experiments, we observe that the \texttt{GreenCourier} scheduler takes $539$ms on average for scoring/assigning carbon-efficient nodes to pods as compared to $515$ms taken by the default K8s scheduler. 

To evaluate the binding latency of our scheduling framework, 
we compare it against a traditional K8s setup that uses a single cluster 
with multiple worker nodes in the same geographical region (\S\ref{sec:expconfig}). 
In the traditional setup,  kubelet is responsible for executing pods on worker nodes and 
managing their state in the K8s control plane (\S\ref{sec:workflow}). However, 
in our scheduling framework which uses Liqo for creating a multi-cluster K8s 
topology, virtual kubelet (VK) (\S\ref{sec:overview}) is used to impersonate an 
independent cluster as a node from the perspective of the management cluster. Figure~\ref{fig:ovearhead} 
shows the binding latencies for the two setups. With \texttt{GreenCourier} and 
Liqo, we observe an average pod binding latency of $8.28$s as compared to $4.53$s with kubelet. 
The increased latency in our framework can be attributed to two reasons. First, due to the resource abstraction added by Liqo via VK, an additional layer of synchronization is required in \texttt{GreenCourier} for communicating state 
information between the management cluster and the provider clusters. Second, frequent 
communication across geographically distributed clusters via the public internet increases the binding latency. In contrast, all communication in the traditional 
setup is done within a virtual private cloud (VPC).

\vspace{-3mm}

\section{Related Work}
\label{sec:relatedwork}

\textbf{Scheduling serverless functions.} While developing strategies for scheduling serverless functions is an active research area~\cite{centralized-ss, fnsched, courier}, no prior work has explored the carbon-aware scheduling of serverless functions across geographically distributed regions. Kaffes et al.~\cite{centralized-ss} propose a centralized scheduler that can schedule functions at a processor core-level granularity, to better address the unique characteristics of serverless workloads. They demonstrate with extensive experiments that their scheduler provides enhanced elasticity to users for function execution and is adaptive to varying function request demands. Suresh et al.~\cite{fnsched} propose \texttt{FnSched}, a scheduling strategy that aims to minimize resource cost for cloud service providers for executing serverless functions. They implement their strategy on OpenWhisk and demonstrate upto $55$\% reduction in host utilization as compared to other strategies, albeit reducing function execution performance. Jindal et. al.~\cite{courier} propose \texttt{Courier}, a framework to load balance scheduling of serverless functions across heterogeneous FaaS deployments such as GCF~\cite{gcloud-functions-2}, AWS Lambda, and OpenWhisk. Courier incorporates an adaptive weighted round-robin scheduling strategy that dynamically adapts the weights of the different FaaS platforms depending on current function performance.
\\
\hspace{-3mm}\textbf{Sustainable scheduling for datacenters.} Most prior work on sustainable scheduling for datacenters has focused on developing energy-aware strategies~\cite{first-fit, earh}, with only a few approaches that target maximizing the use of green energy~\cite{greenslot}. However, no prior work in the literature enables the runtime scheduling of serverless functions across geographically distributed datacenters to reduce carbon emissions. To reduce energy consumption for idle resources, Alahmadi et al.~\cite{first-fit} propose a novel strategy for scheduling, sharing, and migration of VMs in datacenters. For real-time energy-aware scheduling of tasks, Zhu et al.~\cite{earh} propose an energy-aware rolling horizon (EARH) algorithm for scheduling tasks in batches on VMs. In~\cite{greenslot}, Goiri et al. present \texttt{GreenSlot}, a parallel batch job scheduler for a datacenter that uses forecasting techniques to predict the availability of solar energy for scheduling batch jobs.

\vspace{-2mm}

\section{Conclusion and Future Work}
\label{sec:conclusion}

In this paper, we took the first step towards sustainable serverless computing and presented \texttt{GreenCourier}, a scheduling framework that enables runtime carbon-aware scheduling 
of serverless functions across geographically distributed multi-Kubernetes 
clusters. With comprehensive experiments using production function traces, 
we demonstrated that our carbon-aware scheduling strategy leads to 
$13.25$\% average reductions in carbon emissions per function invocation as compared to two other strategies. 
In the future, we plan to extend the management control plane of our framework to make it highly available and disaggregated. Furthermore, we plan to adapt our scheduling strategy to account for service level objectives while scheduling serverless functions.

\vspace{-2mm}

\section{Acknowledgement}
\label{sec:ack}
The research leading to these results was funded by the German Federal Ministry of Education and Research (BMBF) in the scope of the Software Campus program under the grant agreement 01IS17049. Google Cloud credits in this work were provided by the \textit{Google Cloud Research Credits} program with the award number 64c92de5-fb62-4386-8c5b-ff3f480390bb.

\vspace{-3mm}
\bibliographystyle{ACM-Reference-Format}
\bibliography{serverless}

\end{document}